\pdfoutput=1

\documentclass[a4paper,11pt]{article}
\usepackage{pos}

\usepackage{hyperref}
\hypersetup{colorlinks=true, citecolor=blue, linkcolor=blue, urlcolor=blue, pdfborder={0 0 0}, breaklinks=true} 
\renewcommand{\url}[1]{\href{#1}{#1}}
\newcommand{\doi}[1]{\url{https://doi.org/#1}}

\usepackage{comment}
\usepackage{multirow}
\usepackage{datetime2}

\newcommand{\mgamc}{MG5aMC} 

\title{Developments in Performance and Portability for MadGraph5\_aMC@NLO}

\author*[a]{Andrea Valassi}
\author[b]{Taylor Childers}
\author[a]{Laurence Field}
\author[a]{Stefan Hageb\"ock}
\author[b]{Walter Hopkins}
\author[c]{Olivier Mattelaer}
\author[b]{Nathan Nichols}
\author[a]{Stefan Roiser}
\author[a]{David Smith}

\affiliation[a]{IT Department, CERN, Geneva, Switzerland}
\affiliation[b]{Argonne National Laboratory, USA}
\affiliation[c]{Universit\'e Catholique de Louvain, Belgium}

\emailAdd{andrea.valassi@cern.ch}

\abstract{Event generators simulate particle interactions using Monte Carlo techniques, providing the primary connection between experiment and theory in experimental high energy physics. 
These~software packages, which
are the first step in the simulation workflow of collider 
experiments, 
represent approximately 5 to 20\% of the annual WLCG usage for the ATLAS and CMS experiments. 
With computing architectures becoming more heterogeneous, it is important to ensure that these key software frameworks can be run on future systems, large and small. 
In this contribution, recent progress on porting and speeding up the Madgraph5\_aMC@NLO event generator on hybrid architectures, i.e. CPU with GPU accelerators, is discussed.
The main focus of this work has been in the calculation of scattering amplitudes and ``matrix elements'', which is the computational bottleneck of an event generation application.
For physics processes limited to QCD leading order,
the code generation toolkit has been expanded to produce matrix element calculations 
using C++ vector instructions on CPUs 
and using CUDA for NVidia GPUs, 
as well as using Alpaka, Kokkos and SYCL
for multiple CPU and GPU architectures.
Performance is reported in terms of matrix element calculations 
per time on NVidia, Intel, and AMD devices.
The status and outlook for the integration of this work into a production release 
usable by the LHC experiments,
with the same functionalities 
and very similar user interfaces as
the current Fortran version,
is also~described.}

\FullConference{
  {Version 1.0 (20 October 2022)}
  \\\phantom{0}\\\phantom{0}\\
  41st International Conference on High Energy physics - ICHEP2022\\
  6-13 July, 2022\\
  Bologna, Italy
}

\begin{document}

\maketitle

\section{Introduction}

\newcommand{\figflow}[1]{
\begin{figure}[#1]
\vspace*{-5mm}
\begin{center}
\hspace*{0.06\textwidth}\includegraphics[width=0.88\textwidth,clip]{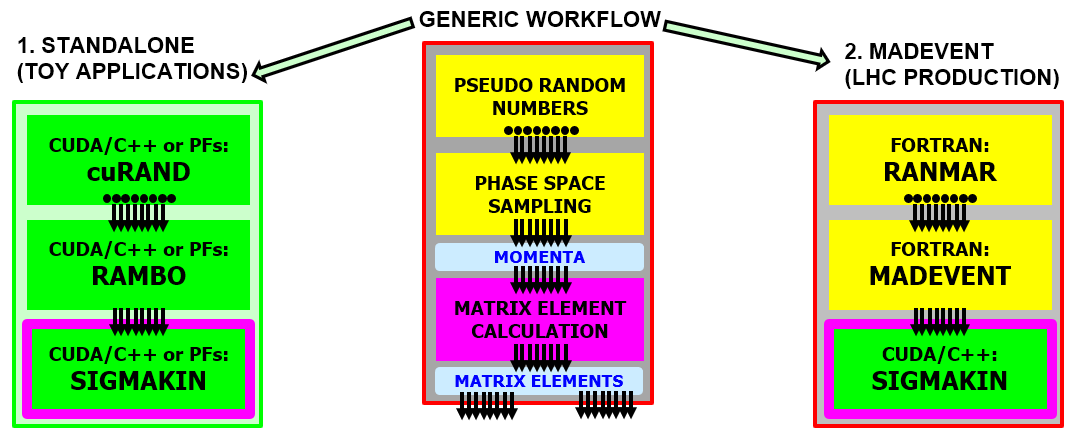}
\end{center}
\vspace*{-6mm}
\caption{Schematic representation
of the internal computational workflow
of a matrix element event generator, 
and of its implementation
for \mgamc\ in the 
applications described in this paper.}
\label{fig:flow}
\vspace*{-4mm}
\end{figure}
}

Physics event generators 
are an essential component 
of the software chain
in HEP experiments.
For the ATLAS and CMS experiments 
at the LHC, they 
are the first step
of simulation workloads and
represent 
5 to 20\% 
of the annual CPU resource budget.
Their internal workflow, where 
Monte Carlo (MC) techniques are used
to draw random samples of events,
for each of which a ``matrix element'' (ME)
is computed independently,
makes these application 
an ideal fit for implementing 
data parallelism with lockstep processing
on GPUs and on vector CPUs.
This is especially interesting because
the computational bottleneck
of matrix element generators,
which can easily take more than 95\%
for complex LHC physics processes,
is the ME calculation.
It is therefore possible 
to obtain large overall speedups 
by efficiently exploiting 
these hardware architectures
for the ME calculation.

In this contribution,
we report on our work
on the reengineering of the Madgraph5\_aMC@NLO
(in the following: \mgamc) 
event generator~\cite{bib:mg5amc},
to port the ME calculation 
from the current Fortran version 
to faster implementations 
using C++ based programming models
on CPUs and GPUs.
This paper, which describes the 
status of our ``madgraph4gpu'' project
at the time of ICHEP in July 2022,
focuses only on the progress achieved since 
our vCHEP presentation in May 2021,
whose proceedings~\cite{bib:vchep2021}
include a more in-depth description 
of our work.
Schematically, the new results
achieved during the last year
can be grouped in three areas:
first, further progress in the ME calculation
in vectorized C++ for CPUs
and CUDA for NVidia GPUs,
including automatic code generation
and new performance 
measurements for complex QCD LO processes
relevant to LHC physics;
second, the parallel development of new
implementations of the ME calculation
using performance portability frameworks (PFs)
such as Alpaka, Kokkos and SYCL,
and initial measurements of their
computing performance on CPUs and GPUs
from different vendors;
third, progress in the integration
of the CUDA/C++ ME calculation
into the existing MadEvent 
framework,
which represents our strategy 
to achieve in a relatively short time
a production release of the software 
that can be used by the LHC experiments
to generate large samples of events
with identical physics output,
but at a fraction 
of the current computational cost.
The structure of this paper is
organized accordingly:
these three specific areas of work
are described in the following 
Sections 2, 3 and 4,
each of which also includes
some hints at the outlook and plans
for our work in that respective area.

Figure~\ref{fig:flow} 
schematically represents
the computational workflow
of a ME 
event~generator~and~its implementation
for \mgamc\ in the two
applications used~in~our~work.
Our ME calculations
in CUDA/C++ and PFs
are developed and optimized using
a ``standalone'' 
application,
where random numbers,
phase space sampling
and other functionalities
outside the ME calculation
are based on
simplified
software components which are fast
and can be executed fully on a GPU,
but~miss many features 
required for production use.
The ``madevent'' application,
conversely, injects the new
ME computational engines
into the existing Fortran 
MadEvent framework~\cite{bib:madevent},
replacing the previous
MEs in Fortran,
but keeping the 
functionalities and user interface
identical to those currently 
used by the LHC experiments, 
except for some additional options 
needed for the parallel event execution.

\figflow{t}

\section{Matrix element calculations
in CUDA and vectorised C++}

\newcommand{\eemumu}{\mbox{$e^+\!e^-\!\!\rightarrow\!\mu^+\!\mu^-$}}
\newcommand{\ggtt}{\mbox{$gg\!\rightarrow\! t\bar{t}$}}
\newcommand{\ggttg}{\mbox{$gg\!\rightarrow\! t\bar{t}g$}}
\newcommand{\ggttgg}{\mbox{$gg\!\rightarrow\! t\bar{t}gg$}}
\newcommand{\ggttggg}{\mbox{$gg\!\rightarrow\! t\bar{t}ggg$}}

\newcommand{\muc}[1]{\multicolumn{2}{c|}{#1}}
\newcommand{\mur}[1]{\multirow{2}{*}{#1}}

\newcommand{\tabtputcpp}[1]{
\begin{table}[#1]
\vspace*{-4mm}
\begin{center}{
\small
\hspace*{-4mm}
\setlength\tabcolsep{5pt} 
\begin{tabular}{|l|c|c|c|c|c|}
\cline{3-6}
\multicolumn{2}{c|}{}& 
\multicolumn{3}{c|}{madevent}& 
\multicolumn{1}{c|}{standalone}\\
\hline
\multirow{2}{*}{\ggttgg} & MEs &
$t_\mathrm{TOT} = t_\mathrm{Mad} + t_\mathrm{MEs}$ &
$N_\mathrm{events} / t_\mathrm{TOT}$ &
\multicolumn{2}{c|}{$N_\mathrm{events} / t_\mathrm{MEs}$}\\
& precision &
[sec] &
[events/sec] &
\multicolumn{2}{c|}{[MEs/sec]}\\
\hline
Fortran(scalar) & double &
38.3 = 2.5 + 35.8 &
2.14E3 (=1.0) &
2.29E3 (=1.0) &
--- \\
\hline
C++/none(scalar) & double&
39.1 = 2.5 + 36.6 &
2.10E3 (x1.0) &
2.24E3 (x1.0) &
2.31E3 \\
C++/sse4(128-bit) & double &
21.1 = 2.5 + 18.6 &
3.89E3 (x1.8) &
4.41E3 (x1.9) &
4.57E3 \\
C++/avx2(256-bit) & double &
10.8 = 2.5 + \hphantom{0}8.3 &
7.60E3 (x3.6) &
9.92E3 (x4.3) &
1.04E4 \\
C++/512y(256-bit) & double &
10.1 = 2.6 + \hphantom{0}7.5 &
8.14E3 (x3.8) &
1.09E4 (x4.8) &
1.17E4 \\
C++/512z(512-bit) & double &
\hphantom{0}7.1 = 2.5 + \hphantom{0}4.5 &
1.16E4 (x5.4) &
1.82E4 (x7.9) &
1.92E4 \\
\hline
C++/none(scalar) & float &
37.8 = 2.5 + 35.3 &
2.17E3 (x1.0) &
2.32E3 (x1.0) &
2.38E3 \\
C++/sse4(128-bit) & float&
11.7 = 2.5 + \hphantom{0}9.3 &
7.00E3 (x3.3) &
8.85E3 (x3.9) &
8.90E3 \\
C++/avx2(256-bit) & float&
\hphantom{0}7.1 = 2.7 + \hphantom{0}4.5 &
1.15E4 (x5.4) &
1.84E4 (x8.1) &
2.01E4 \\
C++/512y(256-bit) & float&
\hphantom{0}6.4 = 2.6 + \hphantom{0}3.8 &
1.28E4 (x6.1) &
2.15E4 (x9.5) &
2.31E4 \\
C++/512z(512-bit) & float&
\hphantom{0}4.8 = 2.5 + \hphantom{0}2.3 &
1.71E4 (x8.1) &
3.65E4 (x16.1) &
4.01E4 \\
\hline
\end{tabular}
}\end{center}
\vspace*{-5mm}
\caption{
Processing times and throughputs 
to generate 81952 \ggttgg\ weighted events,
using the madevent or standalone application.
The ME calculation uses 
Fortran (double precision)
or C++ (double/single precision).
Five different SIMD modes are used in C++.
The fourth column gives throughputs
for the full workflow,
the last two columns give throughputs
for the ME calculation alone.
Results obtained on a single core
of a Juwels Cluster login node 
with Intel Gold 6148 CPUs,
using gcc11.2 builds.}
\label{tab-tputcpp}
\end{table}
}

\newcommand{\tabtputcuda}[1]{
\begin{table}[#1]
\vspace*{-1mm}
\begin{center}{
\small
\hspace*{-4mm}
\setlength\tabcolsep{5pt} 
\begin{tabular}{|l|c|c|c|c|c|c|}
\cline{3-7}
\multicolumn{2}{c|}{}& 
\multicolumn{3}{c|}{madevent}& 
\multicolumn{2}{c|}{standalone}\\
\hline
\multicolumn{2}{|c|}{CUDA grid size}& 
\multicolumn{4}{c|}{8192}& 
\multicolumn{1}{c|}{524288}\\
\cline{1-7}
\multirow{2}{*}{\ggttgg} &
MEs &
$t_\mathrm{TOT} = t_\mathrm{Mad} + t_\mathrm{MEs}$ &
$N_\mathrm{events} / t_\mathrm{TOT}$ &
\multicolumn{3}{c|}{$N_\mathrm{events} / t_\mathrm{MEs}$}\\
& precision &
[sec] &
[events/sec] &
\multicolumn{3}{c|}{[MEs/sec]}\\
\hline
Fortran & double &
58.3 = 5.2 + 53.1 &
1.55E3 (=1.0) &
1.70E3 (=1.0) &
--- &
--- \\
\hline
CUDA & double &
\hphantom{0}6.1 = 5.7 + 0.36 &
1.49E4 (x9.6) &
2.54E5 (x149) &
2.51E5 & 
4.20E5 (x247) \\ 
\hline
CUDA & float &
\hphantom{0}5.7 = 5.4 + 0.24 &
1.59E4 (x10.3) &
3.82E5 (x224) &
3.98E5 & 
8.75E5 (x515) \\ 
\hline
\end{tabular}
}\end{center}
\vspace*{-5mm}
\caption{
Processing times and throughputs 
to generate 90112 \ggttgg\ weighted events,
using the madevent or standalone application.
The ME calculation uses 
Fortran (double precision)
or CUDA (double/single precision).
The fourth column gives throughputs
for the full workflow,
the last three columns give throughputs
for the ME calculation alone.
Results obtained on a single core
of a CERN virtual machine 
with Intel Silver 4216 CPUs
and a dedicated NVidia V100 GPU,
using cuda11.7 and gcc11.2 builds.}
\label{tab-tputcuda}
\vspace*{-3mm}
\end{table}
}

MG5aMC is a code generator, 
written in Python,
which allows the generation 
of the code~for~a chosen physics process
in many languages.
Fortran is 
now the production version,
while~CUDA/C++ is a new back-end
that we developed
based on an earlier C++ version.
At the time of vCHEP2021, 
our engineering~process was made up
of few, very long, development 
cycles 
and only allowed the optimization 
of one physics process at a time.
For this reason, in Ref.~\cite{bib:vchep2021}
we only presented 
results 
for $\eemumu$ collisions.
One major breakthrough achieved
since then
has been the move to a new 
engineering process,
where a Python plugin
for CUDA/C++ code generation
is now itself part of the madgraph4gpu 
repository~\cite{bib:mg4gpu-github},
complementing the upstream MG5aMC 
code-generating
framework~\cite{bib:mg5amc-github}.
While the engineering process 
remains intrinsically iterative,
the development cycles are much shorter 
and more frequent.
Every merge request providing
changes
in the CUDA/C++ code of a physics process
must also include their backport
to the code-generation plugin
and
the re-generation of the CUDA/C++ code
for the full set of physics process 
maintained in the repository
(which presently includes 
SM LO
$\eemumu$, $\ggtt$, $\ggttg$, $\ggttgg$ and $\ggttggg$,
as well as one 
EFT process for comparison).
The main physics process
for new developments, optimizations and tests
is now $\ggttgg$ rather than $\eemumu$:
this is much better not only because 
$\ggttgg$ is of great relevance to LHC physics,
but also because it is a computational task
of much higher arithmetic intensity, 
with limited overhead from memory access and data copy.

\tabtputcpp{t}

\tabtputcuda{t}

The promising speedups 
in the ME calculation that we 
had previously reported for $\eemumu$
are now confirmed
also for $\ggttgg$.
This is 
discussed
in Sec.~\ref{sec:mad},
using
the results listed 
in Tables~\ref{tab-tputcpp} and~\ref{tab-tputcuda}.
For CUDA, speedups in the order 
of several hundreds may be obtained
on an NVidia V100 GPU
when compared to a single 
CPU core:
the exact speedups depend on the specific process
and are somewhat lower 
for $\ggttgg$
than for $\eemumu$, as 
the compute kernel for the former
uses all of the 255 available GPU registers per thread,
while the latter has a lower ``register pressure''.
For C++, it is confirmed that 
our vectorized implementation
of the ME calculation
using event-level data parallelism
makes a maximally efficient use
of CPU vector registers.
One new result
since vCHEP2021 is that we have
now achieved speedups 
of x8 in double precision
and x16 in single precision
on some high-end AVX512 CPUs,
such as the Intel Gold 6148 CPUs
at the J\"ulich HPC center,
using our ``512z'' implementation
which uses the AVX512 instruction set
on 512-bit zmm vector registers.
The reason why 
we do not see these speedups
on other AVX512 CPUs
such as Intel Silver 4216 CPUs
is most likely the absence
of a second FMA unit 
on these CPU models~\cite{bib:agner}.

Another important 
progress
in the CUDA/C++ ME engine
has been its enhancement 
with many features
needed to provide 
the same level of functionality
currently available
in the Fortran version,
such as 
the MadEvent
single-diagram enhancement 
algorithm~\cite{bib:madevent}
for phase space sampling
and 
the ``running''
of the QCD coupling $\alpha_s$.
As described in Sec.~\ref{sec:mad},
our priority for the CUDA/C++ ME engine
is now to provide the 
functionalities and API hooks 
still missing for
its integration into 
the madevent executable. 
Further performance optimizations
are also ongoing, but the focus
has shifted to speeding up the 
overall madevent workflow 
rather than just the ME calculation.

\section{Matrix element calculations
in performance
portability frameworks (PFs)} 

\newcommand{\figpf}[1]{
\begin{figure}[#1]
\vspace*{-5mm}
\begin{center}
\hspace*{0.00\textwidth}\includegraphics[width=1.00\textwidth,clip]{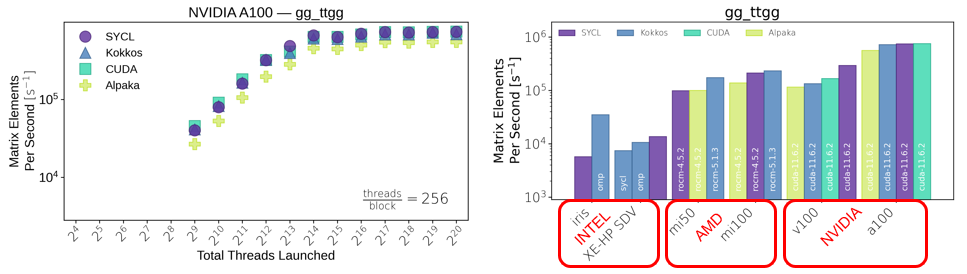}
\end{center}
\vspace*{-5mm}
\caption{
Comparison
of the CUDA, Alpaka, Kokkos and SYCL
ME engines
for $\ggttgg$ 
on many GPUs,
using the standalone application.
Optimal grid sizes
at the throughput plateau
are used in the right plot.
``Xe-HP SDV'' is a Software Development Vehicle
for functional testing only.  
It is currently used at Argonne and at other customer sites 
to prepare their code for future Intel data center GPUs.}
\label{fig:pf}
\vspace*{-4mm}
\end{figure}
}
\figpf{t}

The main interest of abstraction layers,
or performance portability frameworks,
is that~they~allow writing algorithms only once 
with the ability to run on many architectures,
while even including some hardware-specific optimizations.
In our work on MG5aMC, 
the performance of PFs seems
especially promising for GPUs,
as shown in Fig.~\ref{fig:pf}.
While our CUDA/C++ 
version of~MEs~is~presently 
limited to NVidia hardware,
our new three implementations
in Alpaka, Kokkos and SYCL
may also successfully run on AMD GPUs,
and the latter two
on Intel GPUs too.
The other important point is that,
on NVidia V100 and A100 GPUs,
the performances achieved 
by the three new implementations
are comparable to 
our reference CUDA/C++ version,
using four code bases 
that are approximately equivalent
although they miss
some of the features
recently added to CUDA/C++.
On CPUs, one benefit of the three PFs
is that they all provide an out-of-the-box
multi-threading mechanism 
for exploiting all CPU cores,
while in the CUDA/C++ version
this needs to be explicitly added
(an OpenMP prototype existed,
but needs to be rethought and 
has been discontinued for the moment).
Conversely, while the CUDA/C++
implementation is designed
from the ground up to 
efficiently exploit SIMD
through explicit Compiler Vector Extensions,
it is not yet clear whether or how much
our Alpaka, Kokkos and SYCL
ME calculations are benefitting 
from CPU vector registers.

At this point, 
we have not yet decided
which of these 
back-ends will be supported
by MG5aMC in the future.
For the moment, we plan to continue
exploring all of these avenues,
initially~through
further performance optimizations
and tests on both GPUs and CPUs.
We are adding new functionalities 
to the PF implementations, 
but at the same time
we are considering 
adding support~for~AMD GPUs via HIP
or for CPU multi-threading
to the CUDA/C++ version.
Our goal in this context is not only
to release new production versions 
of MG5aMC with multi-GPU support,
but also to provide useful feedback
to the HEP software community
about the usability and performance of PFs.

\vspace*{-2mm}
\section{Integration of 
matrix element calculations
in the MadEvent framework}
\label{sec:mad}
\vspace*{-2mm}

The main issue that we had to address
to inject our CUDA/C++ 
ME engine into the
existing 
Fortran MadEvent framework
was not language interoperability
(which was easily achieved
via a Fortran/C bridge interface),
but 
the fact that the latter 
was designed years ago, 
with serial processing in mind:
a single event was processed
at a time,
and its data properties 
were~allowed~to~be accessed from 
multiple memory locations, including 
COMMON~blocks.
For our goals, it was thus
essential to change the Fortran framework
so that~it~can process
many events in parallel,
turning event properties into large arrays
and redesigning some components
as reentrant functions~with~clearly~defined inputs and outputs.
This is now 
achieved,
but 
MadEvent 
remains an active area 
of development, mainly because 
of two issues.
First, the event property arrays
have a large RAM footprint,
and this 
currently limits the number
of events that can be processed 
in parallel to 8k,
which is a problem on GPUs
because a grid size of 8k threads
provides suboptimal ME throughputs
(as shown in Fig.~\ref{fig:pf},
the throughput plateau in $\ggttgg$
is only reached at around 16k).
Second, even if the ME calculation
takes more than 90\%
of the total CPU time
in the all-Fortran implementation,
the overhead from the rest
quickly becomes the bottleneck
if MEs are computed
one or two order of magnitudes faster.

The points above are 
concretely demonstrated by the 
results shown 
in Tables~\ref{tab-tputcpp} 
and~\ref{tab-tputcuda}.
With respect to the 
ME calculations for $\ggttgg$
in Fortran
(or to the new no-SIMD C++ versions,
which have similar throughputs),
the new vectorized C++ component
achieves a speedup close 
to x8 for doubles 
and x16 for floats
on an Intel Gold 6148.
The speedup 
of the overall workflow,
however, is 
limited 
to approximately x5 and x8,
respectively,
because 
the remaining Fortran components
other than the ME calculation
(sampling algorithm, I/O, merging...), 
now become a significantly large overhead 
in the total 
compute budget. 
The situation is even worse 
for GPUs: on a system
including a Silver 4216 CPU and a V100,
where the scalar part 
of the madevent application
represents 
10\% of the overall CPU time
in the 
all-Fortran version,
Amdahl's law limits the overall speedup
to a factor x10 
(1 over 10\%),
even if that 
of the ME calculation alone
is x149 for doubles 
and x224 for floats.~In~these circumstances
it is not even a problem
that only 8k events per grid are used:
even if ME speedups~of~x247 
and x515
can be achieved with 
larger grid sizes
(as demonstrated by 
the standalone 
application),
Amdahl's law would continue
to limit the overall speedup to x10.
While these 
figures
in the range of x5 to x10
are already very promising,
it is therefore clear that 
further optimizations
on the MadEvent Fortran software, 
or porting further parts onto GPUs, 
could easily 
improve the overall speedups.
Larger overall speedups 
are in any case 
expected
for more complex physics processes,
where the scalar component
uses a lower fraction 
of the overall~compute~budget.

From a functionality point of view,
the integration 
is already very advanced:
in the tests described above,
replacing the Fortran MEs
by their CUDA or C++ equivalents
yields exactly the same cross sections
to within 2E-14 (2E-4)
if doubles (floats) are used,
and it also yields exactly
the same LHE unweighted event files,
except for a reduced precision
in event weights (only for floats),
and for two important differences:
event-by-event helicities 
and leading QCD colors
are still missing
in the LHE files
generated using CUDA/C++ MEs.
These are currently 
the two main pieces
that are still missing
before we can release 
a new, faster, production
version of MG5aMC usable 
by the experiments,
and are therefore 
our main development priority
in the short term.

\small
\section*{Acknowledgements}

We gratefully acknowledge the computing resources 
provided 
by the Joint Laboratory for System Evaluation (JLSE) 
at Argonne National Laboratory. 
This research used 
resources 
of the Argonne Leadership Computing Facility, 
which is a DOE Office of Science User Facility 
under Contract DE-AC02-06CH11357.
We gratefully acknowledge the use 
of computing resources
at CINECA under ISCRA-C project MG5A100
and at the J\"ulich Supercomputing Centre 
at Forschungszentrum J\"ulich
under PRACE-DEV-2022D01-022.


\begin{thebibliography}{99}

\bibitem{bib:mg5amc} 
J. Alwall et al.,
JHEP07(2014)079.
\doi{10.1007/JHEP07(2014)079}

\bibitem{bib:vchep2021}
A. Valassi, S. Roiser, O. Mattelaer, S. Hageboeck,
EPJ Web of Conferences 251, 03045 (2021).
\doi{10.1051/epjconf/202125103045}

\bibitem{bib:madevent}
 F. Maltoni, T. Stelzer, 
 JHEP02(2003)027.
 \doi{10.1088/1126-6708/2003/02/027}
 
\bibitem{bib:mg4gpu-github}
madgraph4gpu project,
{\em github code repository}.
\url{https://github.com/madgraph5/madgraph4gpu}


\bibitem{bib:mg5amc-github}
MadGraph5\_aMC@NLO,
{\em github code repository}.
\url{https://github.com/mg5amcnlo/mg5amcnlo}

\bibitem{bib:agner}
A. Fog,
{\em Agner Fog's blog}.
\url{https://www.agner.org/optimize/blog/read.php?i=962}
 
\end{thebibliography}
\end{document}